\documentclass[aps,prc,twocolumn,superscriptaddress,showpacs,longbibliography,lengthcheck]{revtex4-1}
 
\usepackage{amsmath,amssymb}
\usepackage{graphicx}
\usepackage{color} 
\usepackage{txfonts}

\usepackage{ulem}

\usepackage{xspace}  
\usepackage{multirow}
\usepackage{dcolumn}

\newcommand{\etal}{\textit{et al.}\xspace}

\begin{document}


\title{Moments and Radii of exotic Na and Mg isotopes} 

\newcommand{\aut}{         \affiliation{Department of Physics, The University of Tokyo, 7-3-1 Hongo, Bunkyo, Tokyo 113-0033, Japan}}
\newcommand{\acns}{        \affiliation{Center for Nuclear Study, The University of Tokyo, 7-3-1 Hongo, Bunkyo, Tokyo 113-0033, Japan}}
\newcommand{\ariken}{      \affiliation{RIKEN Nishina Center, 2-1 Hirosawa, Wako, Saitama 351-0198, Japan}}
\newcommand{\ajar}{         \affiliation{Advanced Science Research Center, Japan Atomic Energy Agency, Tokai, Ibaraki 319-1195, Japan} }

 \newcommand{\aemt}{\email{Corresponding author: otsuka@phys.s.u-tokyo.ac.jp}}  
 
\author{Takaharu~Otsuka}   \aemt  \aut \ariken \ajar
\author{Noritaka~Shimizu}   \acns 
\author{Yusuke~Tsunoda}   \acns 

\date{\today}

\begin{abstract}      
The ground-state properties of neutron-rich exotic Na and Mg isotopes with even numbers of neutrons, $N$, are studied up to driplines.  The shell-model calculations with an {\it ab initio} effective nucleon-nucleon interaction reported in  [Tsunoda, Otsuka, Takayanagi {\it et al.}, Nature {\bf 587}, 66 (2020)] are extended to observables such as magnetic dipole and electric quadrupole moments, and charge and matter radii. Good agreements with experimental data are found, and predictions are shown up to driplines.   A prescription to extract the deformation parameters for the eigenstates of Monte Carlo Shell Model is presented, and the obtained values are used to calculate charge and matter radii.  
The increase of these radii from the Droplet model is described as the consequences of the varying deformation of the surface and the growing neutron excitations or occupations in the $pf$ shell, consistently with the dripline mechanism presented in the above reference.  
The neutron skin thickness is shown to be about 0.1 fm for $N$=20, which can be compared to the value for $^{208}$Pb in an $A^{1/3}$ scaling.
The relation of the neutron skin thickness to the electromagnetic moments is discussed for an exotic nucleus, $^{31}$Na. 
\end{abstract}

\maketitle

\section{Introduction \label{intro}}

Exotic nuclei keep providing us with exciting questions and challenges to nuclear physics.  Here, exotic nuclei mean atomic nuclei with unbalanced ratios of the proton number $Z$, and the neutron number $N$, and are transformed to less unbalanced nuclei through $\beta$-decays \cite{gade2008,nakamura2017}.  The destinations of such decays are stable nuclei, where those ratios are closer to unity and the lifetimes are basically infinite.   The exotic nuclei are ``exotic'' in many features as compared to stable nuclei.  While the major subjects there include the formation of the neutron halo \cite{tanihata1996,nakamura2017}, the shell evolution as functions of $Z$ or $N$ \cite{otsuka2020}, {\it etc.}, another one lies in the variation of the surface deformation.
In fact, it has been shown \cite{dripline2020} that the {\it neutron dripline}, meaning the last bound nucleus with the maximum $N$ for a given $Z$ ({\it i.e.} fixed isotope chain), is determined, at least for F, Ne, Na and Mg isotopes, by the mechanism driven by the interplay between the deformation and monopole energies.     
This is in contrast to the other traditional dripline mechanism with the single-particle origin, which is valid, at least, for nuclei with $Z\le$ 8 and yields the neutron halo.   Interestingly enough, these two dripline mechanisms may emerge alternatively as $Z$ increases.      
The neutron driplines not only provide us with intriguing challenges to be explored, but also exhibit crucial conditions or constraints to interdisciplinary studies, for instance, stellar nucleosynthesis.
As the present dripline mechanism may dominate the structure of a number of heavy nuclei,
it is of urgent importance to investigate physical observables of exotic nuclei for the in-depth anatomy of this mechanism. 
 
We focus, in this paper, on the ground-state properties of even-$N$ Na isotopes: magnetic moments, electric quadrupole moments and charge/matter radii.  The radii of even-$N$ Mg isotopes are discussed also.   By extending the calculations of \cite{dripline2020}, we show how existing experimental data can be reproduced, and what are predicted.  The moments and radii are precious probes of various aspects of the ground states.  Besides, from the charge and matter radii, the thickness of the neutron skin is discussed.

\section{Shell Model Framework and the EEdf1 Interaction}

As mentioned in Sec.~\ref{intro}, the theoretical results are obtained by extending the shell-model-based approach of \cite{dripline2020}.  The same Hamiltonian and (single-particle) model space as in  \cite{dripline2020} are taken, but more observables are evaluated, leading us to new insights.  
The model space is composed of all single-particle orbits in the $sd$ and $pf$ shells, on top of the $^{16}$O inert core.  The many-nucleon Hilbert space can be gigantic for this model space, and the Monte Carlo Shell Model (MCSM) \cite{mcsm_rev2001,mcsm_rev2012} is needed to cover all relevant nuclei.  
The present effective nucleon-nucleon ($NN$) interaction was derived, in an {\it ab initio} way from the interaction of the chiral Effective Field Theory \cite{machleidt2011}, and the in-medium renormalization is made  through the extended Krenciglowa-Kuo (EKK) method \cite{takayanagi2011a,takayanagi2011b,tsunoda2014}.  Note that one can 
straightforwardly combine two major shells in the EKK method, whereas this is a challenge in other {\it ab initio} approaches \cite{tsunoda2014,otsuka2020,simonis2017,stroberg2017,miyagi2020}.  The Fujita-Miyazawa three-nucleon force \cite{fujita1957} is incorporated in the manner of \cite{otsuka2010}.  The present effective $NN$ interaction for the $sd$-$pf$ shell was first reported in \cite{tsunoda2017}, and was named EEdf1.

\section{Magnetic and Electric Quadrupole Moments}

\begin{figure}[tb]
  \centering
  \includegraphics[width=8cm]{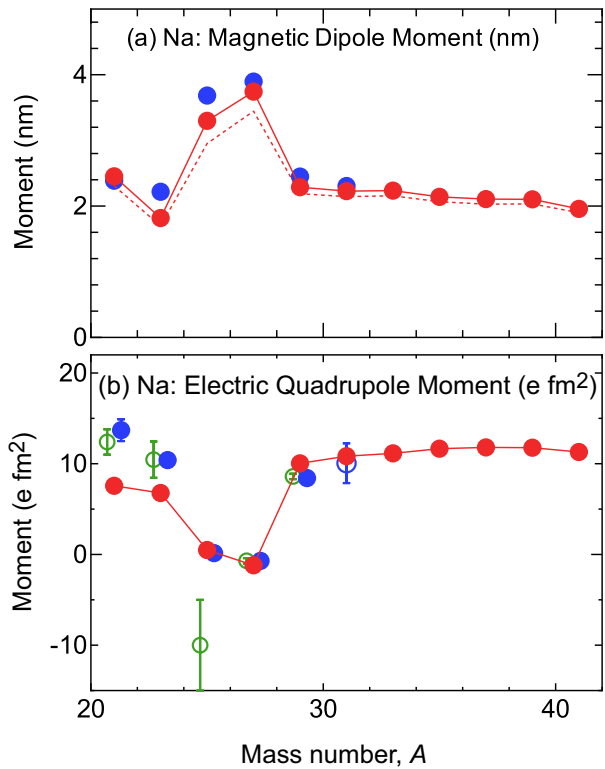}
  \caption{(a) Magnetic and (b) spectroscopic electric quadrupole moments of the ground states of Na isotopes: 3/2$^+_1$ (5/2$^+_1$) states for $^{21,23,29-41}$Na ($^{25,27}$Na). 
Red circles and lines indicate present theoretical results.  In (a), while g-factors take free values for the dashed line, the orbital g-factors 
  include an isovector shift of 0.2 for the solid line.  Blue closed circles in (a) are experimental data from \cite{ensdf}.  In (b), blue closed (open) circles are experimental data from \cite{derydt2013} (\cite{keim1998}), while green symbols are from  \cite{ensdf}.    
  In (b), theoretical results are obtained with the same effective charges as in \cite{tsunoda2017}.
  }   
  \label{fig:moment}  
\end{figure}  

We start with the magnetic moments of the ground states of the Na isotopes, as shown in Fig.~\ref{fig:moment}(a) in comparison to experimental data  \cite{ensdf}.  While the magnetic moments were not touched in \cite{dripline2020}, the present theoretical values are calculated from the same wave functions as in \cite{dripline2020}.  The present work correctly reproduces the ground-state spin/parities ($J^{P}$) except for $^{21,27}$Na, where the lowest  $J^{P}$=5/2$^+$ and 3/2$^+$ states are in the wrong order in the present calculation, but the excitation energies are as small as 30 and 24 keV, respectively.  
Note that the ground states in this paper always refer to those in experiments.  

The magnetic moments obtained with free spin g-factors ($g_s$) and free orbital g-factor ($g_l$) exhibit (dashed line in Fig.~\ref{fig:moment}(a)) reasonable agreement with experimental data.
Because the Schmidt value is 4.8 (0.1) n.m. for $J^{P}$=5/2$^+$ (3/2$^+$), the structures of these nuclei are far from simple single-particle pictures.
The quenching of the spin g-factors is usually needed in shell-model calculations in order to reproduce experimental values  \cite{usd1987,castel_towner_book,towner1987,usd1987,caurier2005,honma2009}.   For instance, a standard value of the quenching is 0.85 \cite{usd1987}.  For Na isotopes, a quenching factor 0.9 is used in \cite{utsuno2004} where the SDPF-M shell-model interaction was taken for the model space composed of the $sd$ shell and the lower part of the $pf$ shell.  The present work fully activates the $sd$ and $pf$ shells and the EEdf1 interaction is derived in an {\it ab initio} way as stated above. The quenching of spin g-factor is considered to be mainly due to two-particle-two-hole (2p2h) excitations across shell gaps \cite{ichimura1965,shimizu1974,towner1983,arima1987,towner1987}.  Such 2p2h excitations are not included in shell-model calculations with one major shell, where the quenching of spin g-factors are then needed.  In contrast, because the $sd$ and $pf$ shells are fully included and the $NN$ interaction is adequate for the mixing between the two shells, it is natural that the room for the spin g-factor quenching becomes narrower.

Figure~\ref{fig:moment}(a) also shows that the agreement is improved by introducing an isovector shift of 0.2 to the orbital g-factors ($g_l$ = 1.2 for protons and -0.2 for neutrons).  In \cite{utsuno2004}, a similar shift of 0.15 was taken.  Such a shift is considered to originate in the meson exchange effect directly on the magnetic process \cite{castel_towner_book,towner1987}.  This effect is independent of the configuration mixing, and can occur irrespectively of the spin g-factor quenching.
The M1 operator can, in principle, be derived similarly to the effective $NN$ interaction.  It is of interest to see what consequences may arise in such future studies, and we hope this work provides some hints.  
 
The magnetic moment of $^{31}$Na may have particular importance in the present study.  
Because of $N$=20, if no neutrons were excited from the $sd$ to the $pf$ shell, this nucleus would have a closed neutron shell, implying that its structure is close to the single-particle
picture and that its ground-state magnetic moment is basically given by the Schmidt value. Indeed, the calculation only with the proton $sd$ shell and the USD interaction \cite{usd1987} yields the 5/2$^+$ ground state and its magnetic moment, 4.4 n.m., in disagreement with experiments.  
We shall come back to this point later in the relation to the neutron skin.
  
Figure~\ref{fig:moment} (b) displays the spectroscopic electric quadrupole moments of the ground states of Na isotopes.  The effective charges are the same as in \cite{tsunoda2017}: 1.25e (0.25e) for protons (neutrons) with $e$ being the unit charge.  A salient agreement with experiment \cite{derydt2013,keim1998,ensdf} is seen, particularly for $^{29,31}$Na isotopes.  
Note that we multiply absolute magnitudes shown in \cite{derydt2013} by the signs from the present calculation.
Combined with the nice description of the $B$(E2;2$^+_1 \rightarrow 0^+_1$) values depicted in \cite{tsunoda2017}, the present description of the quadrupole deformation appears to be quite appropriate.  The strong deformation of the $^{31}$Na ground state was pointed out by a deformed Hartree-Fock  
calculation \cite{campi1975}, which became a landmark in the study of exotic nuclei.  
The $N$=20 closed shell yields a tiny quadrupole moment of the ground state of $^{31}$Na.
 
\section{Charge Radius and Deformation Parameters}

Having the features described so far, we move on to the charge radius.  
The main part of the radius is accounted for by the Droplet Model, where a sphere with an equal density is assumed.   On top of that, 
it is known \cite{bohr_mottelson_book2} that the radius changes if the shape of the nuclear surface is deformed from a sphere.
We restrict ourselves, in this work, to the quadrupole deformation as the major source of
this change.
The mean-square charge radius is usually written as (see, for instance, \cite{otten1989,campbell2016}):
\begin{equation}
\label{eq:rad}
\langle r^2 \rangle_{ch} \, = \, \langle r^2 \rangle_{\rm DM} \{ 1 + (5/4\pi) \, \beta_2^2\},
\end{equation}
where $\beta_2$ denotes the deformation parameter, and $\langle r^2 \rangle_{\rm DM}$ means the Droplet Model value given by
\begin{equation}
\label{eq:droplet}
\langle r^2 \rangle_{\rm DM} \, = \, (3/5) \, (R_0 A ^{1/3})^2.
\end{equation}
Here, $R_0$ is a parameter and $A$ stands for the mass number, $A=Z+N$.  We use $R_0$= 1.28(fm) in the present work.

\begin{figure*}[tb]
  \centering
  \includegraphics[width=17.8cm]{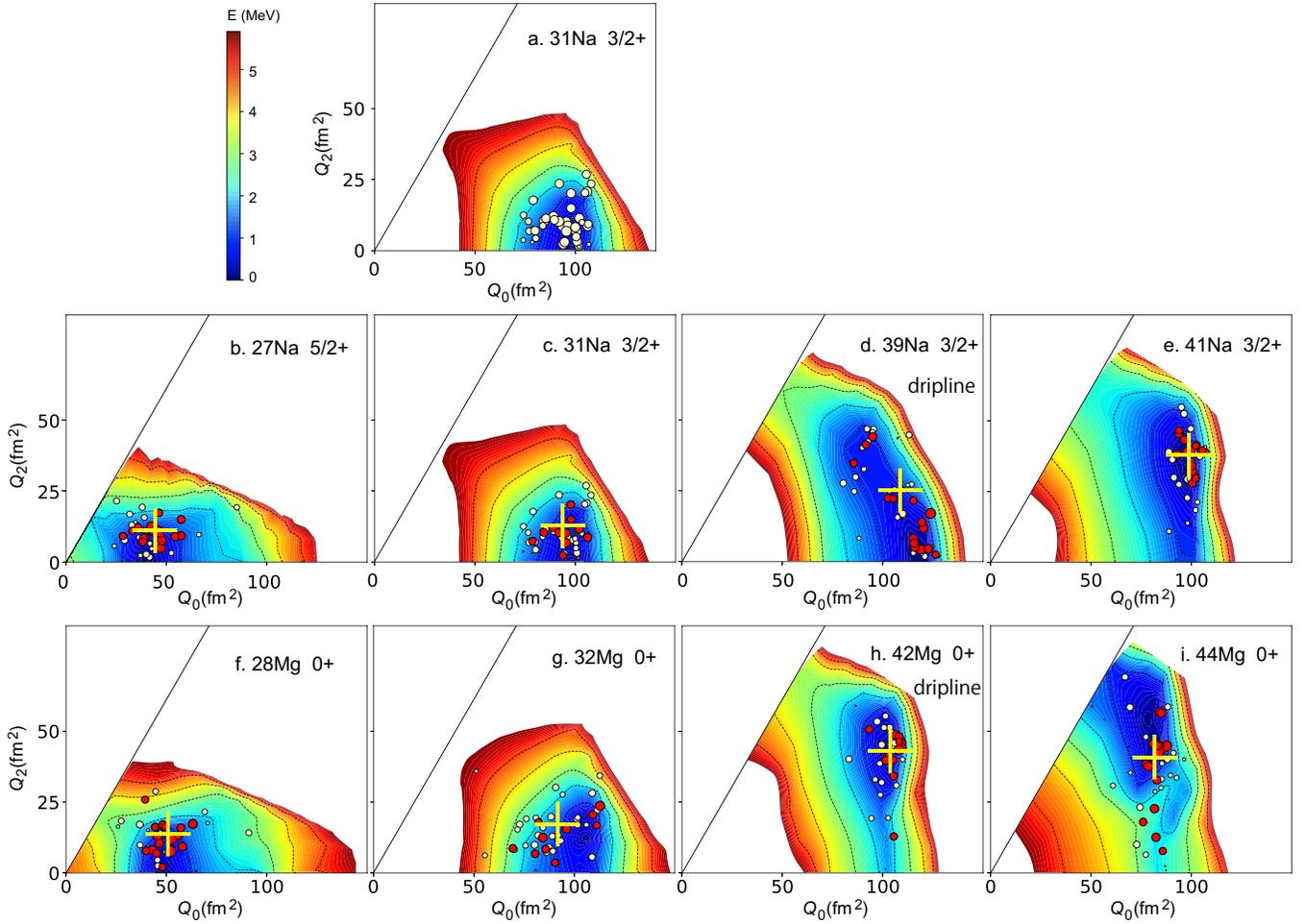}
  \caption{T-plot for the ground states of (a,b,c,d,e) Na and (f,g,h,i) Mg isotopes.
 (a) The usual T-plot.  (b-i) A new analysis (see text) is introduced with red circles implying larger overlap probabilities in eq.~(\ref{eq:xi_2}) ($>$0.03) with the MCSM ground states.  The extracted $Q_0$ and $Q_2$ values are shown by crosses.
Probable (Na) \cite{ahn2019} and predicted (Mg) \cite{dripline2020} driplines are indicated. 
  }   
  \label{fig:tplot}  
\end{figure*}  

We now need the values of the deformation parameter, $\beta_2$ as input to eq.~(\ref{eq:rad}). 
This is not straightforward, and we take advantage of the T-plot \cite{ytsunoda2014,otsuka2016} of the MCSM \cite{mcsm_rev2001,mcsm_rev2012}.

The quadrupole(-deformed) shape can be expressed in terms of so-called deformation parameters, $\beta_2$ and $\gamma$, which imply, respectively, the magnitude of the deformation and the proportion of the ellipsoid axes \cite{bohr_mottelson_book2,ring_schuck_book}.  For instance, $\gamma=0^{\circ}$ (60$^{\circ}$) corresponds to a prolate (oblate) shape, while $0^{\circ}<\gamma<60^{\circ}$ is generally referred to as triaxial shapes.  The potential energy surface (PES) is drawn by the constrained Hartree-Fock calculation for the Hamiltonian being used. The constraints 
are given by quadrupole matrix elements in the intrinsic (or body-fixed) frame, $Q_0$ and $Q_2$ in the standard notation \cite{ring_schuck_book}.  
The values of ($Q_0$, $Q_2$) are related to those of ($\beta_2$, $\gamma$) by the formula \cite{utsuno2015}
\begin{equation}
\label{eq:beta}
\beta_2 \, = \, f_{scale} \sqrt{5/16\pi} (4\pi/3 R_0^2 A^{5/3}) \sqrt{(Q_0)^2+2(Q_2)^2}
\end{equation}
and $\gamma = \arctan{(\sqrt{2} Q_2 / Q_0)}$, where $f_{scale}$ is the rescaling factor for the isoscalar quadrupole operator.  The E2 operator is $(e+e’_p)Q_p + e’_n Q_n$ with $Q_{p,n}$ denoting proton or neutron quadrupole operator and $e’_{p,n}$ being their induced charges due to in-medium effects.  Additional terms in the isoscalar case are assumed to be $e’_n Q_p + (e+e’_p) Q_n$ obtained by exchanging coefficients of the proton and neutron terms of the E2 operator.  
This holds exactly for $N=Z$ nuclei with a perfect charge symmetry, and is considered to be a good approximation otherwise.  
We thus obtain $f_{scale}=(e+e'_p+e'_n)/e$, and eq.~(\ref{eq:beta}) is shown to work well in a variety of studies including \cite{leoni2017,morales2017,marsh2018,sels2019,otsuka2019}.  


\begin{figure}[bt]
  \centering
    \includegraphics[width=8.5cm]{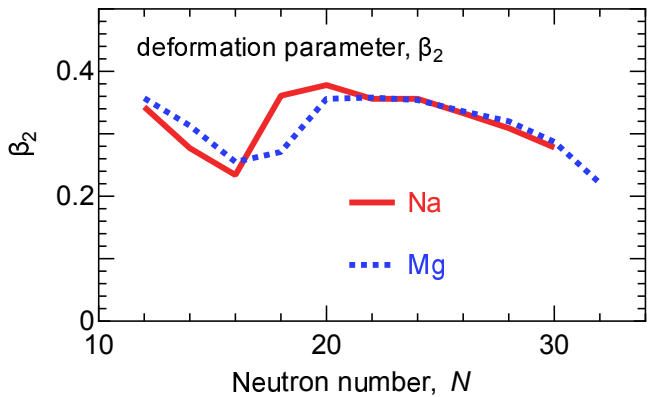}    
  \caption{ Derived $\beta_2$ values 
of Na (red solid lines) and Mg (blue dashed lines) isotopes as functions of the neutron number, $N$.    
  }   
  \label{fig:beta_skin}  
\end{figure}  

The eigenstates of the Hamiltonian are obtained by the MCSM: the eigenstate is expanded by MCSM basis vectors, $\{ \mathcal{P} \phi_i \} \, (i=1,...,n)$, where $\phi_i$'s are deformed Slater determinants and $\mathcal{P}$ is the projection operator onto the designated angular momentum and parity, {\it e.g.} $J^\pi$=3/2$^+$ for $^{31}$Na.  For each $\phi_i$, intrinsic quadrupole moments, ($Q_0$, $Q_2$), are calculated.  We plot individual $\phi_i$ on the PES according to such ($Q_0$, $Q_2$) values. Namely, ($Q_0$, $Q_2$) are used as partial coordinates of $\phi_i$. The importance of $\phi_i$ to the $l$-th eigenstate, $\xi_l$, can be represented by the overlap probability,
$|\langle \xi_l \,|\, \{\mathcal{P} \phi_i \} \rangle |^2 / \langle \{\mathcal{P} \phi_i \} \,|\, \{\mathcal{P} \phi_i \} \rangle$.  Thus, $\phi_i$ can be plotted as a circle at the ``coordinate'' ($Q_0$, $Q_2$) on the PES, with the area proportional to this overlap probability.  This scheme is called the T-plot \cite{tsunoda2014,otsuka2016}, and has been used in many works, for instance \cite{togashi2016,kremer2016,leoni2017,morales2017,togashi2018,marsh2018,sels2019,ichikawa2019,otsuka2019}.

Figure~\ref{fig:tplot} (a) exemplifies the T-plot for the 3/2$^+_{1}$ ground state of $^{31}$Na. 
One sees the distribution of $\phi_i$'s on the PES: human eyes may observe the mean values of $Q_0$ and $Q_2$ around 95 and 10(fm$^2$), respectively.  As $\beta_2$ can be calculated from these values, it is of importance to extract appropriate values of $Q_0$ and $Q_2$ in a certain process free of human eyes.   

\section{Extraction of Deformation Parameters from T-plot}

We present a prescription inspired by the approach proposed in \cite{rodriguez2010}.  
We start with the norm matrix $\mathcal{N}$ defined by matrix elements
$\mathcal{N}_{ij} = \langle \{\mathcal{P} \phi_i \} \,|\, \{\mathcal{P} \phi_j \} \rangle$ with $i, j = 1,..,n$ with $n$ being the number of MCSM basis vectors.   Noting that $\mathcal{N}$ is hermitian, we diagonalize it with the k-th eigenvalue, $n_k$, and the i-th component of its eigenvector, $g_{ki}$.  As the matrix $\bigl( g_{ki} \bigr)$ is unitary, the states defined by
\begin{equation}
\label{eq:psi}
\psi_k \,=\, \frac{1}{\sqrt{n_k}} \, \sum_i \, g_{ki} \, \{\mathcal{P} \phi_i \} \,, \,\,\, k=1, ..., n,
\end{equation}
form a set of orthonormal state vectors.   The eigenstate of the Hamiltonian is expanded as
\begin{equation}
\label{eq:xi}
\xi_l \,=\, \sum_k \, f_{lk} \, \psi_k \,, 
\end{equation}
where $f_{lk}$'s are amplitudes, and the matrix  $\bigl( f_{lk} \bigr) $ is unitary.
We here restrict ourselves to the ground state denoted by $l$=1.
In practice, one of the states $\psi_k$'s exhibits a large overlap with $\xi_1$.
For the sake of clarity, $k$=1 means the $\psi_k$ with the largest overlap hereafter.
We note that $|f_{11}| \gtrsim$ 0.9 holds in the examples being discussed.    
In such cases, replacing $n_k$ by a $k$-independent appropriate constant $n_0 \approx n_1$, we obtain 
\begin{equation}
\label{eq:xi_2}
\xi_1 \,\approx\, \frac{1}{\sqrt{n_0}} \, \sum_{i} \, \Bigl\{\sum_{k} \,  f_{1k}\, g_{ki}  \Bigr\} \, \{\mathcal{P} \phi_i \} \, . 
\end{equation}
Once we have eq.~(\ref{eq:xi_2}), as $\sum_{k} \,  f_{lk} g_{ki}$ forms a unitary matrix, the state  $\{\mathcal{P} \phi_i \}$ is interpreted to be included in the eigenstate $\xi_1$ with the probability $|\sum_{k} \,  f_{1k} g_{ki}|^2$.  
A mean value of $Q_0^2$, is taken as $\sum_i {Q_0}_i^2   |\sum_{k} \,  f_{lk} g_{ki}|^2$, where ${Q_0}_i$ is $Q_0$ of $\phi_i$.  The mean value of $Q_2^2$ is treated similarly.  From them, we obtain $\beta_2^2$, and input it to eq.~(\ref{eq:rad}), which gives us the radius. Note that cross terms due to non-orthogonalites are absorbed, to a large extent, by $g_{ki}$ and $n_k$, as this is the case in the examples shown in Fig.~\ref{fig:tplot}.   
We present an intuitive sketch of the present process by employing a simple example (without projections) comprised of two basis vectors $\phi_1 = u + v$ and $\phi_2 = u - v,$ where $|u| \approx 1, |v| \ll $1, $|u|^2 + |v|^2=1$, and $u$ and $v$ are orthogonal. Two vectors $\phi_1$ and $\phi_2$ are not orthogonal, but are independent.   We assume that $\phi_1$ and $\phi_2$ are tractable ({\it e.g.} Slater determinants), whereas $u$ and $v$ are rather complicated to be treated explicitly.  
If the state of interest happens to be $\xi =u$, $\xi$ is expressed as $\xi = (1/2) \{\phi_1 + \phi_2\}$.  
After some mathematical manipulations, for an operator $\mathcal{O}$, the expectation value can be approximated as $\langle \xi |\mathcal{O}| \xi \rangle \,\approx \, ( \langle \phi_1 |\mathcal{O}| \phi_1 \rangle \,+\, \langle \phi_2 |\mathcal{O}| \phi_2 \rangle)/2$, which corresponds to the aforementioned argument.   
If $\langle v |\mathcal{O}| u \rangle \approx \langle v | u \rangle \langle u |\mathcal{O}| u \rangle  =0$ holds, the above approximation of $\langle \xi |\mathcal{O}| \xi \rangle$ becomes better.  In the present study, through more complex actual processes, the representative value of ${Q_{0,2}}$ are extracted. 
 
Figure~\ref{fig:tplot} (b-e) show, respectively, the T-plot for the ground states of $^{27,31,39,41}$Na, with $^{39}$Na probably at the dripline \cite{ahn2019}.  The area of T-plot circles represent the overlap probabilities $|\sum_{k} \,  f_{1k} g_{ki}|^2$.  The circles yielding larger values ($>$0.03) are shown in red.  Panels (a) and (c) display different patterns, but the differences are small.  The extracted mean values of $Q_0$ and $Q_2$ are indicated by the positions of the crosses.  These positions turn out to be remarkably close to what are suggested by human sight.  

Figure~\ref{fig:tplot} (b) displays that T-plot circles are concentrated in the domain 
around ($Q_0, Q_2) \approx$ (40,10)(fm$^2$) for $^{27}$Na.
In panels (c,d), such concentration moves to the regions around ($Q_0, Q_2) \approx$ (95,15) and (110,25)(fm$^2$) for $^{31}$Na and $^{39}$Na, respectively.  Panel (e) exhibits that the T-plot circles substantially move backward for $^{41}$Na, implying that $^{41}$Na gains less binding energy from the deformation than $^{39}$Na, locating the dripline at $^{39}$Na. 
Detailed discussions on this type of underlying mechanism for the dripline are found in \cite{dripline2020}.   Figure~\ref{fig:tplot} (f-i) show similar features for Mg isotopes, 
where the dripline was predicted at $^{42}$Mg in \cite{dripline2020}.

Figure~\ref{fig:tplot} as a whole shows that the $\gamma$ value remains finite and never approaches 0$^\circ$ (a prolate shape with the axial symmetry).  In other words, triaxial shapes are favored by all the nuclei there.  To be more in detail, the $\gamma$ value is 15-20$^\circ$ around $N$=20, and becomes larger even somewhat beyond $\gamma$=30$^\circ$ in more exotic isotopes, in both Na and Mg chains.  The relation between the triaxiality and the dripline is of great interest, and the triaxiality appears to be one of the essential degrees of freedom in predicting/describing driplines.

Figure~\ref{fig:beta_skin} depicts $\beta_2$ values thus calculated as a function of $N$.  
The onset of deformation is seen already at $^{29}$Na consistently with Fig.~\ref{fig:moment}(b) \cite{tsunoda2017,himpe2008}.
It is of interest that for $N \ge$ 20, the $\beta_2$ values are almost the same between Na and Mg for the same $N$.  We find that $N$=16 behaves like a magic number.  The $\beta_2$ value remains large from $N$=18 or 20 upto driplines. This is consistent wih the shell evolution scenario \cite{otsuka2010a,otsuka2020}.   A sustained strong deformation appears also in the merged-island-of-inversion picture \cite{caurier2014}, in contrast to the original picture of the island of inversion \cite{warburton1990}.   Experimentally, as a reference, $\beta_2$ is reported as 0.43 (3) for the $2^+_1$ state of $^{32}$Mg \cite{takeuchi2009,ensdf}, lying just above the present value in Fig.~\ref{fig:beta_skin}.  The $\beta_2$ value is 0.501 (30) or 0.0484 (38) from the $B(E2;2^+_1 \rightarrow 0^+_1)$ value of $^{32}$Mg \cite{BE2}. 
Because the present calculation reproduces the electric quadrupole moment for $^{29,31}$Na (see Fig.\ref{fig:moment}) and also because the basic trend in Fig.~\ref{fig:beta_skin} is consistent with the compilation in \cite{BE2}, these (minor) discrepancies may be better understood in future studies including those among different probes.

\begin{figure}[tb]
  \centering
  \includegraphics[width=7.5cm]{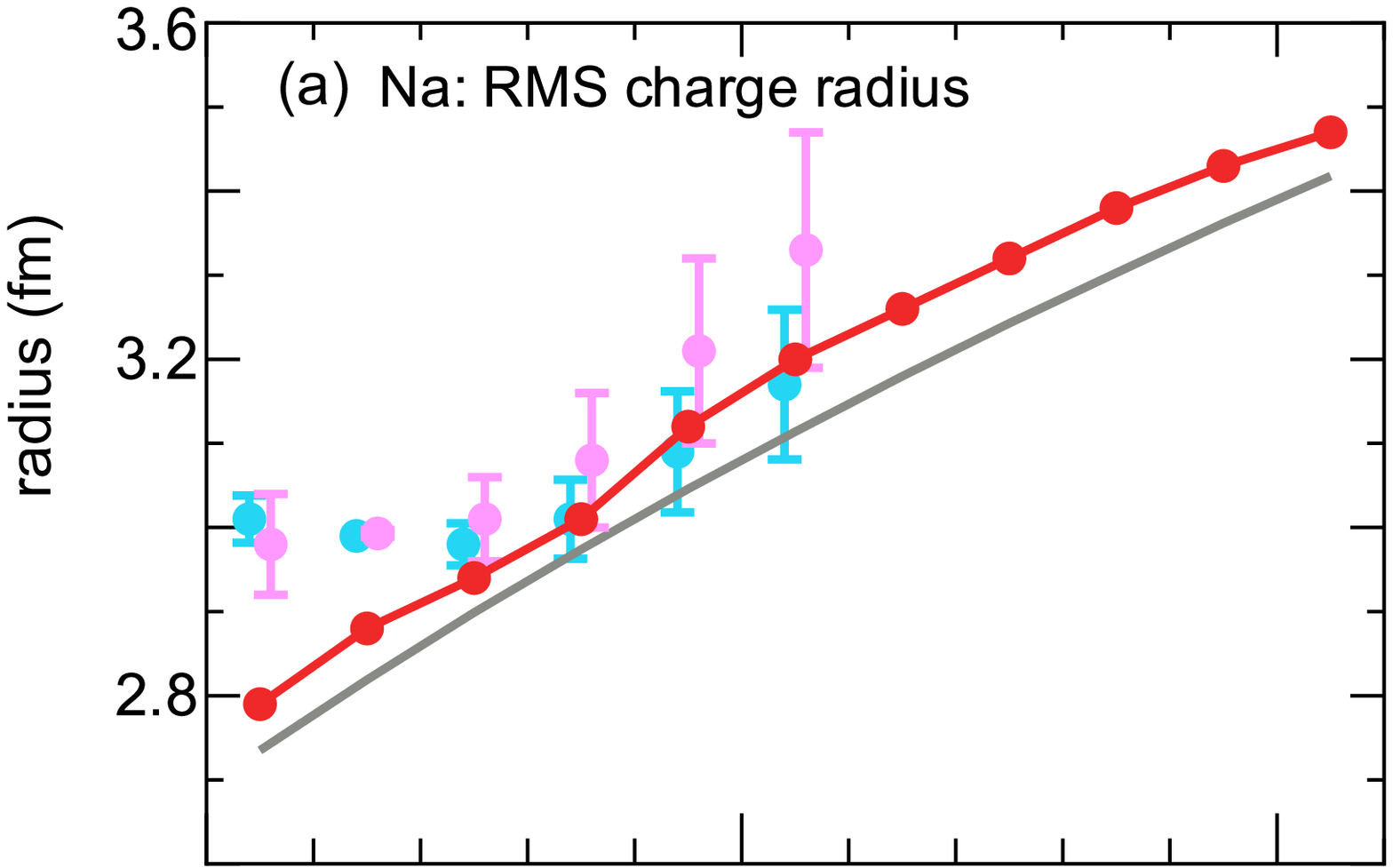}
    \includegraphics[width=7.5cm]{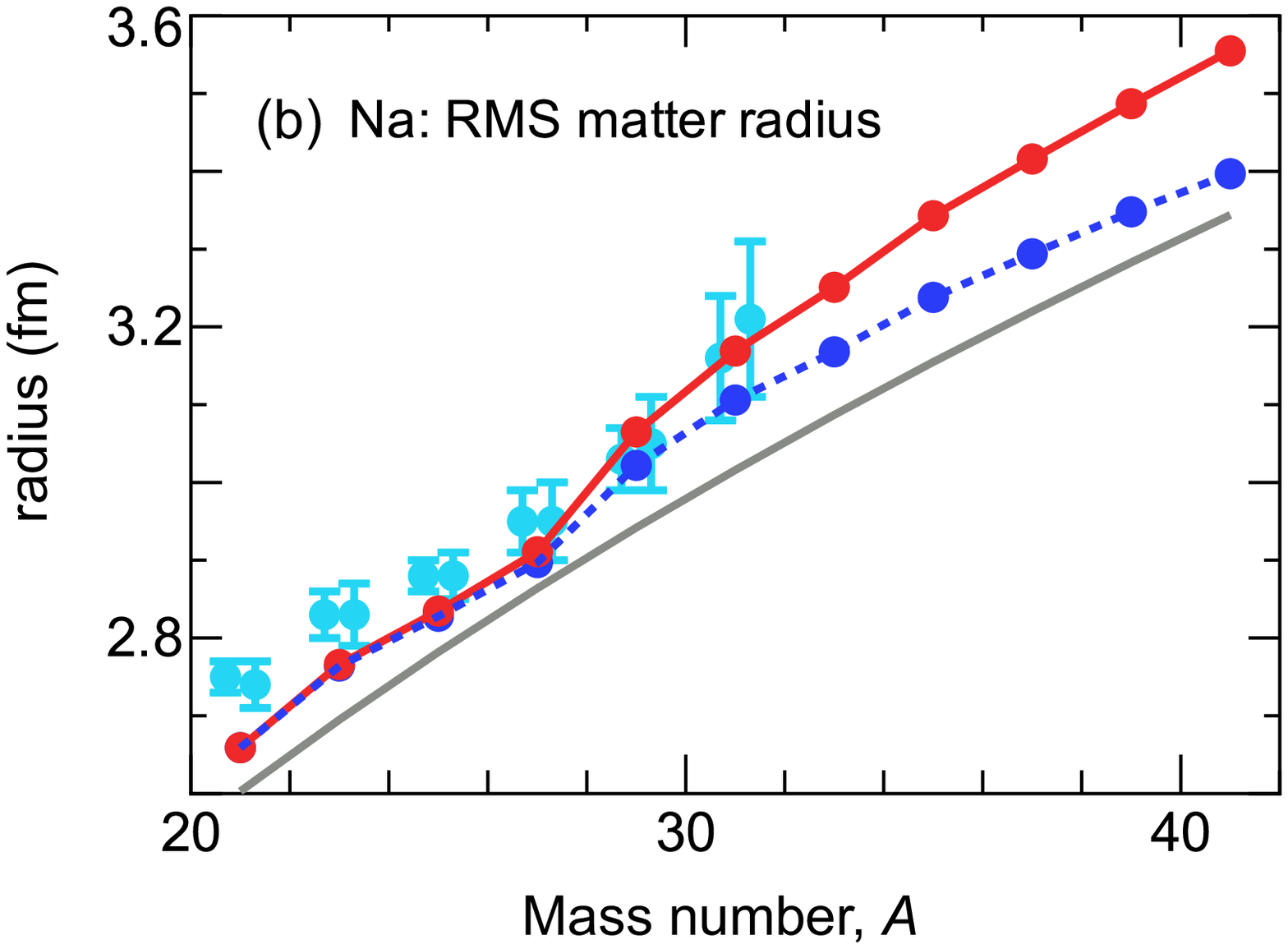}
  \caption{
Charge and matter RMS radii of Na isotopes (red and dark blue symbols connected by lines) in comparison to experimental ones (other symbols). Blue and red symbols include the deformation effect, while the latter in (b) contains effects of neutrons in the $pf$ shell.  Grey solid lines represent the spherical drop model.  Regarding experimental data, 
(a) Light blue (purple) symbols are from \cite{adndt2013} (\cite{ohayon2021}).  
(b) Light blue symbols denote data obtained by two slightly different analyses in \cite{suzuki1998}.  }   
\label{fig:radius}  
\end{figure}  

\begin{figure}[tb]
  \centering
    \includegraphics[width=7.5cm]{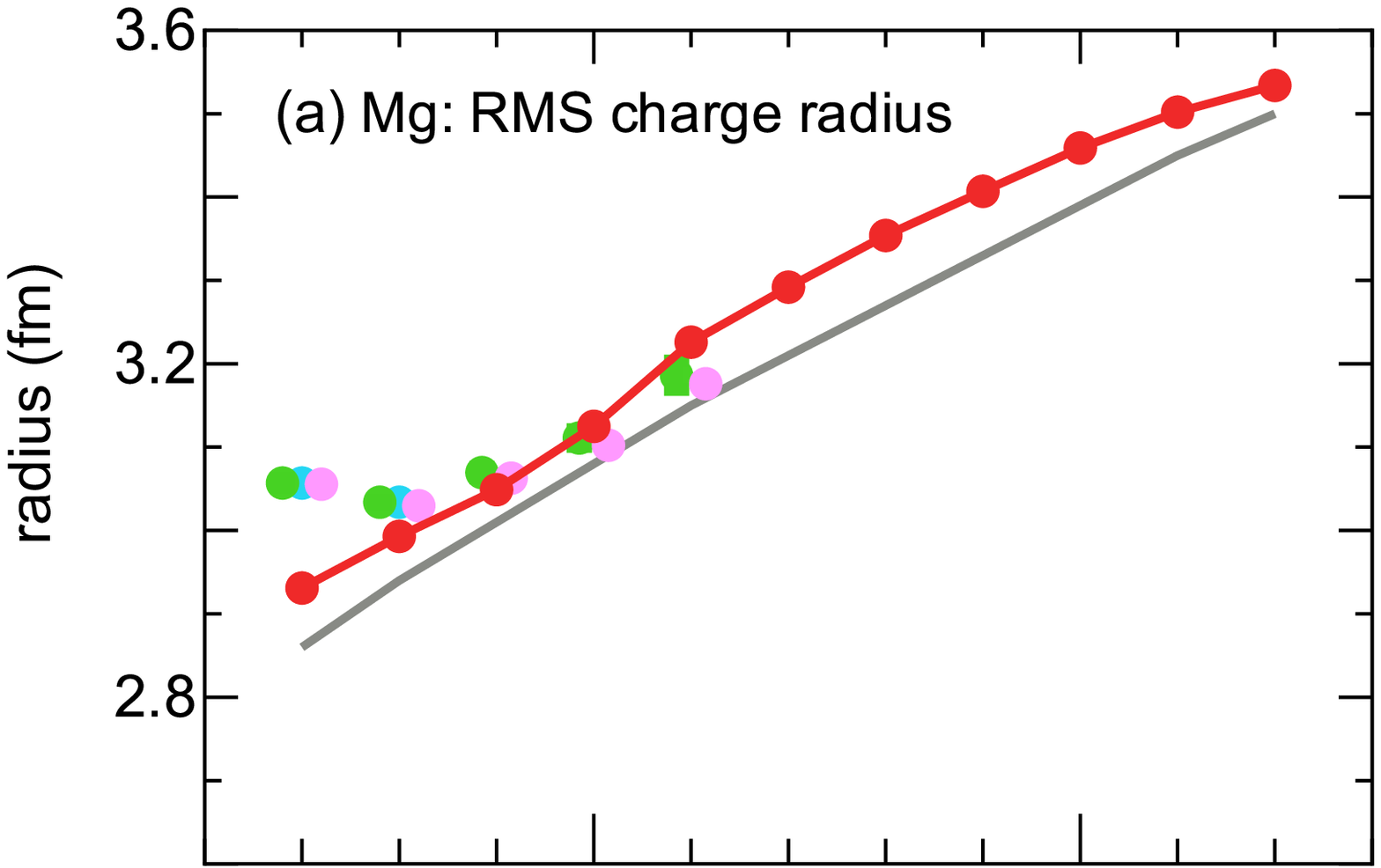}
     \includegraphics[width=7.5cm]{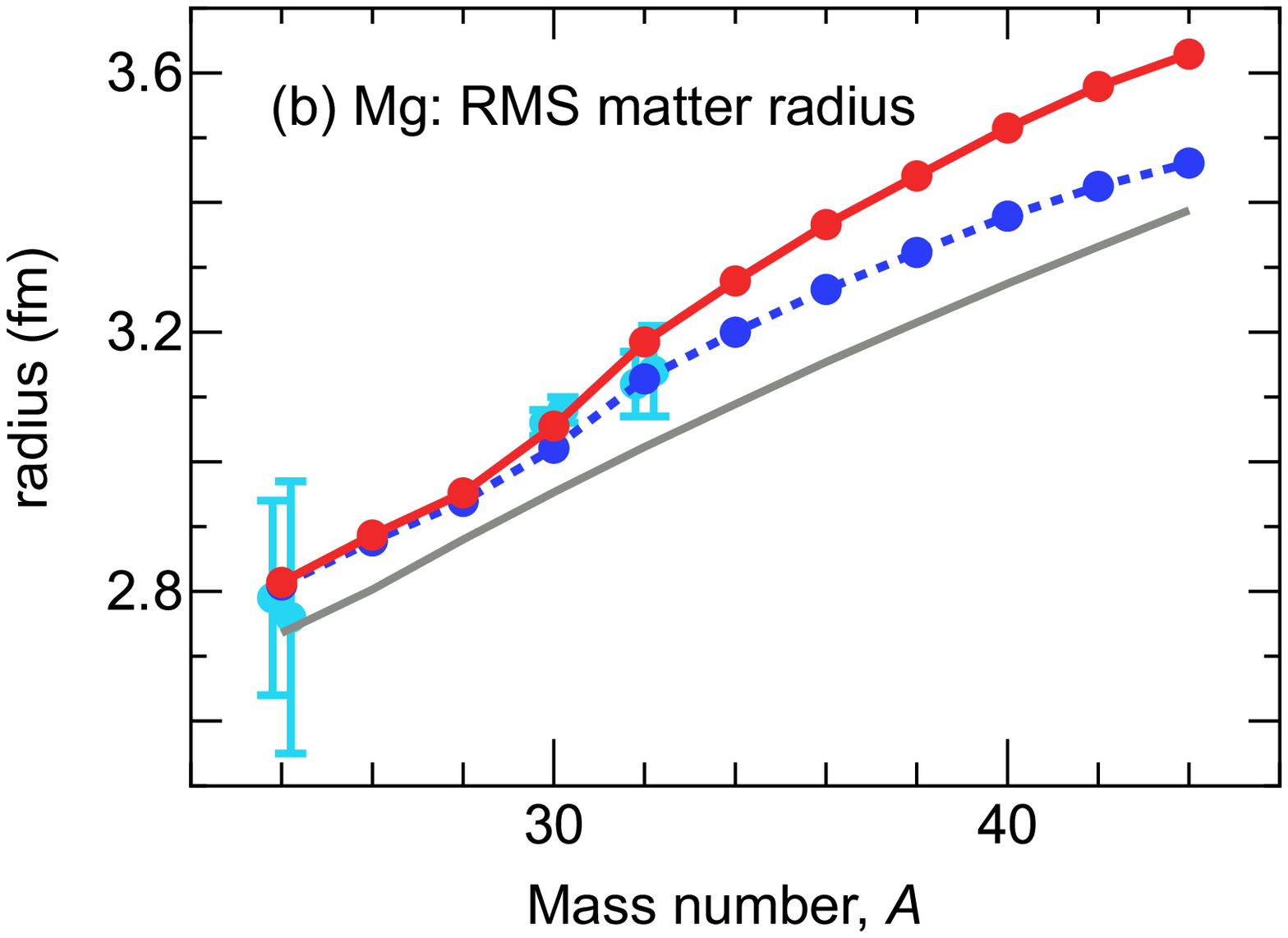}
  \caption{
Charge and matter RMS radii of Mg isotopes (red and dark blue symbols connected by lines) in comparison to experimental ones (other symbols). See the caption of Fig.~\ref{fig:radius}.  Regarding experimental data, 
(a) Light green, light blue, and light purple symbols are taken, respectively, from \cite{yordanov2012}, \cite{adndt2013} and  \cite{ohayon2021}.  (b) Light blue symbols denote data obtained by two slightly different analyses in \cite{suzuki1998}.  }   
\label{fig:radius2}  
\end{figure}  

\section{Charge and Matter Radii}

We now calculate the charge radii by eq.~(\ref{eq:rad}) with the extracted $\beta_2$ values.
Figure~\ref{fig:radius}(a) shows the calculated root-mean-square (RMS) charge radii in a reasonable agreement with experimental data for $^{25-31}$Na, a compilation \cite{adndt2013} and a recent work \cite{ohayon2021}.  We stress that 
there is no adjustment for this agreement, and that 
the nice reproduction of the experimental data, particularly the increase from $N$=14 or 16 to $N$=20, appear to be very promising for further studies on more exotic isotopes up to the dripline.  
We note that the radius increase due to deformation is caused by 2$\hbar\omega$, 4$\hbar\omega$ or higher mixings among  
single-nucleon states ($a\, la$ Nilsson-model-type mixing).
The Coupled-Cluster calculation in \cite{ohayon2021} yields a smaller RMS charge radius $\le$ 3.1 fm for $^{31}$Na, which can be a natural outcome if a strong deformation exceeds the limit of the calculation.  

The observed charge radii are substantially larger than the present values for $^{21,23}$Na, most likely due to proton-halo-type phenomena due to the weaker binding.
While this is an interesting subject, it is outside the scope of the present work. 


The charge radius can be related to the point-proton (mean square) radius, $\langle r^2 \rangle_{pp}$ by \cite{friar1997,abe2021},
\begin{equation}
\label{eq:chg_point}
\langle r^2 \rangle_{ch} \, = \, \langle r^2 \rangle_{pp} \, + \, \langle R^2_p \rangle \, + \, 
\langle R^2_n \rangle \, \frac{N}{Z} + \frac{3 \hbar^4}{4m^2_p},
\end{equation}
where $\langle R^2_p \rangle$ ($\langle R^2_n \rangle$) is the squared charge radius of a proton (neutron), and $m_p$ denotes the proton mass.  The actual values are $\langle R^2_p \rangle$ = 0.77(fm$^2$) and $\langle R^2_n \rangle$ = -0.11(fm$^2$) \cite{adndt2013}.  

The matter (mean square) radius is defined \cite{suzuki1998} as   
$\langle r^2 \rangle_{pm} = (Z \langle r^2 \rangle_{pp} + N \langle r^2 \rangle_{pn}) /\, A$, where $\langle r^2 \rangle_{pn}$ is the point neutron (mean square) radius.
To start with, we assume $\langle r^2 \rangle_{pn}$=$\langle r^2 \rangle_{pp}$.  
This is favored by the strong proton-neutron attraction, and is consistent with the Droplet model, where the radius, $\propto$ $A^{1/3}$, is the same for all pairs of ($Z$, $N$) with a fixed $A$=$Z+N$.   The harmonic oscillator potential (HOP) gives an interpretation that protons and neutrons exhibit the same value of $\langle r^2 \rangle$ if they are in the same HOP shell.
With this modeling $\langle r^2 \rangle_{pn}$=$\langle r^2 \rangle_{pp}$, 
the RMS matter radius is calculated, and is compared with 
experimental data \cite{suzuki1998} (see dark blue symbols in Fig.~\ref{fig:radius} (b)).   A good agreement is seen for $^{29}$Na, but the calculated value is around the lower edge of the error bar for $^{31}$Na.  
The relation between $\langle r^2 \rangle_{pn}$ and $\langle r^2 \rangle_{pp}$ may change once $N \gg Z$ holds and excess neutrons are in the higher shell. 
The difference of $\langle r^2 \rangle_{pn}$ from $\langle r^2 \rangle_{pp}$ is estimated, with the HOP wave functions, as $\hbar n_{pf}/(m\omega)$ where $m$ is the nucleon mass, $\omega$ is the HOP parameter, and $n_{pf}$ means the number of neutrons in the $pf$ shell, provided by the shell model calculations.   
The calculated RMS matter radii are shown by red solid line in Fig.~\ref{fig:radius} (b), yielding an improved agreement with experiment for $^{31}$Na \cite{suzuki1998} due to neutron excitations to the $pf$ shell.  As protons are hardly excited to the $pf$ shell in the present cases, this mechanism is relevant only to neutrons. 
We note that the radius increase due to deformation is caused by 2$\hbar\omega$ or higher excitations of single nucleons, and therefore is different in character from 
the present increase due to neutron excitations or occupations in the $pf$ shell.  It is not appropriate to directly calculate $\langle r^2 \rangle$ values from the present shell model wave functions, because of substantial renormalization contained in the EKK calculations.  This task remains for the future. 

The present results on Mg isotopes are shown in Fig.~\ref{fig:radius2} (a,b), in comparison to experimental data \cite{adndt2013,ohayon2021,yordanov2012,suzuki1998}.
The RMS charge and matter radii exhibit reasonable agreements with experimental ones, while error bars are generally shorter than for Na isotopes.

\section{Neutron Skin Thickness}

We now touch on the neutron skin \cite{myers1985,krasznahorkay1991, fukunishi1993}.
As discussed above, the neutron radius $\langle r^2 \rangle_{pn}$ gains extra increase in neutron-rich isotopes due to excitations or occupations in the $pf$ shell, meaning that the density distribution of neutrons spreads outward and the neutron skin is formed.  
The neutron-skin thickness is defined as $\sqrt{\langle r^2 \rangle_{pn}} \,-\sqrt{\langle r^2 \rangle_{pp}}$, and its calculated value turned out to be 0.10 fm for $^{31}$Na, while it is as small as 0.02 fm for $^{27}$Na.  The predicted thickness is 0.19 fm for $^{39}$Na (dripline).   Similar values are obtained for Mg isotopes as a function of $N$.  

We here recall that the neutron skin thickness is related, in the present approach, to the number of neutrons in the $pf$ shell.  As discussed above, this value is also crucial to the magnetic and electric quadrupole moments of $^{31}$Na, suggesting their interesting relation to the neutron skin thickness.

The neutron skin thickness is experimentally evaluated as 0.156-0.283 fm for $^{208}$Pb \cite{tamii2011,adhikari2021}.  Because of 
(31/208)$^{1/3}$ = 0.53, the present value is on the simple $A^{1/3}$ scaling 
despite apparent differences between $^{31}$Na and $^{208}$Pb.  This is a very intriguing subject for future.  In \cite{suzuki1995}, the neutron skin thickness was shown to be larger ({\it e.g.} $\sim$0.2-0.4 fm for $^{31}$Na, likely due to smaller proton RMS radii then available).   

\section{Summary}

In summary, we discussed ground-state properties of exotic Na and Mg isotopes, with the {\it ab initio} EEdf1 interaction applied in the $sd$+$pf$ model space.   
The calculated results depict agreements with experiments, and suggest that the structure evolves as a function of $N$ in those exotic nuclei in the way presented in \cite{dripline2020}.
We showed how the charge and matter radii can be calculated within the shell model including deformation effects, thanks to the well-known formula, eq.~(\ref{eq:rad}).  The estimation of $\beta_2$ values was made, first by human sight and more objectively by utilizing the prescription inspired by \cite{rodriguez2010}, while the obtained $\beta_2$ values do not differ too much between these approaches.  These developments, some of which may be scrutinized more in the future, enable us to calculate and predict the radii as consequences of various correlations emerging from nuclear forces acting on many valence nucleons.  The neutron skin thickness is assessed, as it is of importance in its relevance to the nuclear compressibility (see for instance  \cite{tamii2011}).  
The radii and moments are vital measures of the ground states, and are shown to be accessible by shell-model approaches.  Their further clarifications are of great importance in the explorations into the {\it terra incognita} including driplines.   
\\

\section*{Acknowledgements}
The authors acknowledge the partial support from MEXT as ``Priority Issue on post-K computer'' (Elucidation of the Fundamental Laws and Evolution of the Universe) (hp160211, hp170230, hp180179, hp190160), the partial support from MEXT as ``Program for Promoting Researches on the Supercomputer Fugaku'' (Simulation for basic science: from fundamental laws of particles to creation of nuclei) (JPMXP1020200105) (hp200130, hp210165),  and JICFuS.  This work was supported in part by MEXT KAKENHI Grant Nos. JP19H05145 and JP21H00117.  
The authors thank Prof. T. Suzuki for useful helps on magnetic properties, 
Prof. G. Neyens for suggesting some references, Prof. R. F. Garcia Ruiz for providing valuable information, and Dr. F. O. A. P. Gustafsson for useful help.  



\end{document}